\documentclass[preprint]{aastex}
\begin{document}
\title{Towards a Spectral Technique for Determining Material Geometry Around Evolved Stars: Application to HD 179821}
\author{J. Nordhaus\altaffilmark{1,2}} 

\and

\author{I. Minchev\altaffilmark{1}, B. Sargent\altaffilmark{1}, W. Forrest\altaffilmark{1}, E. G. Blackman\altaffilmark{1,2}} \author{O. De Marco\altaffilmark{3}, J. Kastner\altaffilmark{4}, B. Balick\altaffilmark{5}, A. Frank\altaffilmark{1,2}}

\affil{1. Dept. of Physics and Astronomy, Univ. of Rochester,
    Rochester, NY 14627}\affil{2. Laboratory for Laser Energetics, Univ. of Rochester, Rochester, NY 14623}\affil{3. Dept. of Astrophysics, American Museum of Natural History, New York, NY 10024}\affil{4. Center for Imaging Science, Rochester Institute of Technology, Rochester, NY 14623}\affil{5. Dept. of Astronomy, Univ. of Washington, Seattle, WA 98195}
\begin{abstract}
HD 179821 is an evolved star of unknown progenitor mass range (either post-Asymptotic Giant Branch or post-Red Supergiant) exhibiting a double peaked spectral energy distribution (SED) with a sharp rise from $\sim8-20$ $\mu$m.  Such features have been associated with ejected dust shells or inwardly truncated circumstellar discs.  In order to compare SEDs from both systems, we employ a spherically symmetric radiative transfer code and compare it to a radiative, inwardly truncated disc code.  As a case study, we model the broad-band SED of HD 179821 using both codes.  Shortward of 40 $\mu$m, we find that both models produce equivalent fits to the data.  However, longward of 40 $\mu$m, the radial density distribution and corresponding broad range of disc temperatures produce excess emission above our spherically symmetric solutions and the observations.  For HD 179821, our best fit consists of a $T_{eff}=7000$ K central source characterized by $\tau_V\sim1.95$ and surrounded by a radiatively driven, spherically symmetric dust shell.  The extinction of the central source reddens the broad-band colours so that they resemble a $T_{eff}=5750$ K photosphere.  We believe that HD 179821 contains a hotter central star than previously thought.  Our results provide an initial step towards a technique to distinguish geometric differences from spectral modeling.

%Our technique highlights the importance of sub-millimeter and millimeter observations.
\end{abstract}

\section{Introduction}

HD 179821 (IRAS 19114+0002; V1427 Aql; SAO 124414; AFGL 2423) is an evolved star surrounded by gas and dust ejected during a phase of mass-loss.  The luminosity of this object is undetermined as the Hipparcos parallax measurement ([0.18 $\pm$ 1.12 mas]; \citealt{1997A&A...323L..49P}) allows for any distance greater than 1 kpc.  HD 179821 is either a close post-Asymptotic Giant Branch (post-AGB) star ($D=1$ kpc, $M_i\sim3-4$ $M_\odot$) or a distant, massive ($D=5-6$ kpc, $M_i\sim20-30$ $M_\odot$) post-Red Supergiant (post-RSG).  The post-AGB phase of stellar evolution is a short ($\sim10^3$ yrs) period in which initially intermediate-mass main sequence stars ($<8$ $M_\odot$) transition from the Asymptotic Giant Branch to the planetary nebula (PN) phase.  In contrast, the progenitors of Red Supergiants are massive main-sequence stars ($>8$ $M_\odot$) evolving toward the Wolf-Rayet stage and may end their lives as supernovae.  For a short review on the post-RSG status of HD 179821, see \cite{2008arXiv0801.2315O}.  In both phases, mass is ejected into the circumstellar environment.  Distinguishing between these classes of objects can be difficult.

In this paper, we model the broad-band SED of HD 179821 using two distinct radiative transfer codes.  The first code, ``DUSTY", computes radiative transfer through a spherically symmetric shell with canonical density profile $\rho(r)\propto r^{-2}$ (\citealt{DUSTY}).  The second code computes radiative transfer through an axisymmetric, flared disc with canonical density profile $\rho(r)\propto r^{-3/2}$ (\citealt{2001ApJ...553..321D,2005ApJ...621..461D}).  While we believe the mid-IR and radio imaging of HD 179821 indicates a roughly spherical nebula, the point of this paper is to determine whether spectral modeling, through the use of distinct radiative transfer codes, can constrain geometric features of evolved star nebulae.  Thus, HD 179821 serves as a test case for this investigation.

We model HD 179821 for both nebular geometries and investigate the degeneracies between the SED model fits.  In addition, we present and compare the ISO spectrum of HD 179821 to the post-AGB object HD 161796 and post-RSG object IRC +10420.  All three objects display a strong mid-IR excess and exhibit steep increases in their SED's beyond 8 microns.  In Section 2, we discuss the enigmatic nature of HD 179821 in both the post-RSG and post-AGB scenarios.  In Section 3, we describe our synthetic photospheres and extinction corrections.  In Section 4, we present our SED model fits from the radiative transfer disc code.  In Section 5, we present our model fits from DUSTY and identify wavelength regions in which degeneracies occur between the two codes.

For intermediate mass stars below 8 $M_\odot$, the late stages of stellar evolution are characterized by transitions from  quasi-spherical mass-losing AGB stars to complex aspherical post-AGB objects.  Many post-AGB objects feature toroidal density enhancements, bipolar jets and an array of other non-spherically symmetric features, the origin of which is unknown (\citealt{2002ARA&A..40..439B}).  In addition, most post-AGB objects display a large momentum excess above what would be supplied by radiation pressure (\citealt{2001A&A...377..868B}).  While the origin of the additional source of momentum is unclear, a binary companion is an attractive candidate as energy and angular momentum can transfer from the secondary to the primary (\citealt{JN2006}).  This hypothesis is supported by recent observational (\citealt{2004ApJ...602L..93D,2007arXiv0709.1508D}) and theoretical (\citealt{2006ApJ...650..916M,2006ApJ...645L..57S}) efforts suggesting that a significant fraction of PNe may be descendants of interacting binaries.  Binary companions can influence mass loss in many ways.  A common envelope phase, even for low-mass companions, can lead to equatorial outflows, poloidal outflows and the formation of discs (\citealt{1999ApJ...524..952R}, \citealt{JN2006}, \citealt{JN2007}, \citealt{2007arXiv0707.3792N}).  For wider binaries, low-mass companions can induce spiral waves and convert amorphous dust to crystalline dust via annealing (\citealt{2007arXiv0709.2292E}).

For massive stars, the end stages of stellar evolution are poorly understood.  The post-red supergiant (post-RSG) phase is among the most luminous and uncertain epochs of post-main sequence stellar evolution.  Like their intermediate mass conterparts, post-RSG's are thought to be ejecting a substantial portion of their mass (\citealt{2006A&A...456..549D}).  However, a massive circumstellar envelope has only been detected (scattered light, IR emission, molecular line emission) around the post-RSG IRC +10420, thus making a comparison with HD 179821 difficult (\citealt{1995ApJ...452..833K}).  

Mass-loss may be consistent with a radiation driven spherically symmetric outflow (\citealt{2007A&A...465..457C}).  However, if HD 179821 does indeed display a momentum excess in its ejected nebula (\citealt{2001A&A...377..868B}), then it is not unreasonable for a binary companion to have influenced the outflow and produced asymmetries.  Interestingly, the WFPC2 images of \citealt{2000ApJ...528..861U} show collimated bipolar structures emerging from the dust shell.  These may be remnant jets propagating through the envelope.  There is also evidence for slight clumpy regions in both OH maser and mid-IR emission (\citealt{2001MNRAS.328..301G,1995ApJ...452..833K}).  A density asymmetry may also be present as the $^{13}$CO line profiles are asymmetric (\citealt{1992A&A...257..701B}).

The spectral energy distribution of HD 179821 exhibits a double-peaked shape, indicative of a stellar photospheric component and ejected dusty component.  The spectral energy distribution (SED) is consistent with photospheric emission to $\sim8$ $\mu$m at which point there is a steep increase until the second peak at $\sim$ 25 $\mu$m.   The overall SED of HD 179821 is thus remarkably similar to that of the transitional young stellar object CoKu Tau/4 (see Fig. 2 of \citealt{2005ApJ...621..461D}), which has been well modeled as a dusty disc with an inner hole.  The sharp, interior wall of the disc is illuminated by the central source and produces the excess infrared emission.  HD 179821 also exhibits an evacuated interior region between the central object and hence, the system might be modeled similarly to CoKu Tau/4.  However, the dust envelope of HD 179821 could also be a detached spherical shell and past modeling efforts have treated it as such (\citealt{2002Ap&SS.281..751S,2007A&A...462..637B,2007A&A...465..457C}).  

\section{HD 179821: post-AGB or red supergiant?}

HD 179821 (galactic coordinates ($l=$ 35.62$^\circ$, $b=$-4.96$^\circ$) is usually classified as a G5 star (\citealt{1989ApJ...346..265H}; see \S 2.1).  It exhibits large mm-wave CO line widths ($\sim70$ km/s) centered on the local standard of rest velocity $V_{LSR}\sim100$ km/s (\citealt{1986ApJ...311..345Z}; \citealt{1992A&A...257..701B}).  OH masers have been detected (\citealt{1989ApJ...344..350L}; \citealt{2001MNRAS.328..301G}) and the presence of a $\sim10$ $\mu$m silicate feature confirm that this object is oxygen-rich (O/C $\simeq$ 2.6; \citealt{1999AJ....117.1834R}).

On the whole, the chemical composition of HD 179821 differs from that of an average F supergiant and is consistent with a post-AGB star (\citealt{1999AJ....117.1834R}).  In particular, an overabundance of s-process elements suggest extra-mixing during the AGB phase and is further supported by a low isotopic $^{12}C/^{13}C\leq$ 5 ratio consistent with deep mixing in a post-AGB object (\citealt{1995ApJ...453L..41C,2001A&A...367..826J}).  Underabundances of carbon and the s-process element zirconium suggest that HD 179821 left the AGB before the third dredge up, consistent with its O-rich properties.  The presence of an active photochemistry and the production of HCO$^+$ are detected in the outflows of many post-AGB objects including HD 179821 (\citealt{2001A&A...367..826J}).  Currently HCO$^+$ has only been detected in one red supergiant star (VY CMa; \citealt{2007Natur.447.1094Z}).

%Another signature of post-AGB objects is the presence of an active photochemistry produced through shock dissipation in interacting outflows or from UV photons emitted from the increasingly hot central object.  A by-product of these ionization fronts is the production of HCO$^+$ which has been detected in the outflows of many post-AGB objects including HD 179821 (\citealt{2001A&A...367..826J}).  Currently HCO$^+$ has only been detected in one red supergiant star (VY CMa; \citealt{2007Natur.447.1094Z}).

In general, the chemical composition is similar to that of high latitude, hot post-AGB stars and in particular bears a strong resemblance to HD 161796 (\citealt{2000A&A...359..138T,1999AJ....117.1834R}).  The ISO SWS spectrum of HD 179821 and broad-band SED also appear very similar to that of the post-AGB star HD161796 ($=$ IRAS 17436+5003 $=$ V814 Her $=$ SAO 30548) (see Fig. \ref{geometry}).  Also shown in Fig. \ref{geometry} is the ISO SWS spectrum of the post-RSG IRC +10420 ($=$V1302 Aql $=$ IRAS 19244+1115).  All three spectra show a substantial infrared excess, however there are differences between HD 161796 and IRC +10420.  The similarities in the ISO spectra for HD 179821 and HD 161796 suggest that both systems have undergone similar mass-loss histories.  

Based on the kinematic distance of $\sim6$ kpc, HD 179821 is a post-RSG.  However, this result should be viewed with caution as the methods used to derive the kinematic distance are not valid for the galactic coordinates of HD 179821 (\citealt{2001A&A...367..826J}).  If in fact HD 179821 is 6 kpc away, then its latitude indicates a position of 500 pc above the galactic plane.  This location is approximately 5-6 times the scale height for supergiants (\citealt{1987pgim.book.....S}).

The strength of the OI triplet line and the distance determined from the interstellar Na I D1 and D2 components imply that the object is very luminous (\citealt{1999AJ....117.1834R}).  Na and Na I emission at 2.21 $\mu m$ is similar to that detected in known supergiants (\citealt{1994ApJ...420..783H}) and  abundances of some molecules are 40 times lower than the averages in post-AGB objects (\citealt{2007A&A...471..551Q}).  CO outflow observations also support the post-RSG hypothesis (\citealt{2001A&A...377..868B}).  In particular, they found a quasi-spherical, unique component to the wind with a large outflow velocity of $V_{exp}= 33-35$ km/s.  This is higher than most post-AGB spherical wind outflow velocities ($\sim10 - 15$ km/s) (\citealt{1995ApJ...452..833K}).  

In short, whether this object is a post-AGB star evolving toward a PNe phase or a post-RSG star whose fate is to explode as a SNe remains unresolved.

%4.  Large outflow velocity, $V_{exp}= 33-35$ km/s are rare in post-AGB objects whose typical outflow velocities are ~$10 - 15$ km/s.  Also $V_{LSR}\sim 95-100$ km/s.  There is a large uncertainty in the Hipparcos parallax measurement ($0.18\pm 1.12$ mas).  Thus, it should not be used to infer distance.  

%2.  High-resolution optical spectra and LTE atmospheric model techniques yield photospheric parameters: $T_{eff}=6750\pm 250$ K, log $g=0.5\pm 0.25$ and $\left[M/H\right] = 0.0$ indicating a higher spectral type then previously published G5 result (\citealt{1999AJ....117.1834R}).  

\section{Photospheric Models and Extinction}
Because of uncertainty in line-of-sight visual and IR extinction (interstellar $+$ circumstellar), the reduction in  magnitude of the HD 179821 spectrum at a given wavelength is unknown.  The inferred extinction must at least be enough to make the maxima of the double peaks equal on the de-reddened SED.  Adopted values of $A_v$ have ranged from 1.8 to 4 (\citealt{1994A&A...285..551V}; \citealt{1989ApJ...346..265H}; \citealt{2002Ap&SS.281..751S}; \citealt{1995ApJ...452..314H}).

The choice of $A_v$ can greatly influence the inferred geometry of the system.  The flux ratio of the peak of the infrared excess to that of the peak of the photosphere, $\alpha$, determines the fraction of the central radiation intercepted by circumstellar material.  If $\alpha\sim1$, then the circumstellar material is intercepting most of the radiation from the central object, corresponding to a spherical shell geometry.  If, however, $\alpha<1$, the geometry can be torus- or disc-like with radiation from the central source escaping through bipolar cavities.  Increasing $A_v$ raises the corrected photospheric emission relative to the longer infrared wavelengths, which are only marginally affected by extinction.  The photometric data is presented in Tables 1 and 2.  

Since the extinction is unknown, the temperature of the central source may be uncertain.  Low-resolution spectra have classified HD 179821 as a luminous G-supergiant (G Ia \citealt{1981AJ.....86..553B}; G4 0-Ia \citealt{1983BICDS..24...19K}; G5 Ia \citealt{1984mscs.book.....B}; G5 Ia \citealt{1989ApJ...346..265H}) while the spectral type inferred from high-resolution spectra indicates $T_{eff}=6800$ K (\citealt{1996MNRAS.282.1171Z}), $T_{eff}=6750$ K (\citealt{1999AJ....117.1834R}) corresponding to an F star.  In addition, \citealt{2001AstL...27..156A} conclude from their 1999 spectra that HD 179821 was an F star in the 1990's.  If the extinction is purely interstellar, then a synthetic photosphere of $T_{eff}=5750$ K (G spectral type) would correspond to $A_v=2$ while $T_{eff}=7000$ K (F spectral type) would correspond to $A_v=3$ (\citealt{Kurucz}).  For both radiative transfer models, we assume the central source is a $T_{eff}=7000$ K star.   However if the surrounding nebula is optically thick, the broad-band colours are reddened and the central source obscured.

%We have de-redend the photometric fluxes and ISO spectra using $A_v$ values of 2 and 3.  For each case, we fit the visual and infrared photometry by determining the best fitting synthetic photosphere (\citealt{Kurucz}).  For $A_v=2$, this corresponds to a model photosphere of $T_{eff}=5750$ K (G spectral type) while for $A_v=3$ this corresponds to a model photosphere of $T_{eff}=7000$ K (F spectral type).

%Low-resolution spectra have classified HD 179821 as a luminous G-supergiant (G Ia \citealt{1981AJ.....86..553B}; G4 0-Ia \citealt{1983BICDS..24...19K}; G5 Ia \citealt{1984mscs.book.....B}; G5 Ia \citealt{1989ApJ...346..265H}) while the spectral type inferred from high-resolution spectra indicates $T_{eff}=6800$ K (\citealt{1996MNRAS.282.1171Z}), $T_{eff}=6750$ K (\citealt{1999AJ....117.1834R}) corresponding to an F star.  In addition, \citealt{2001AstL...27..156A} conclude from their 1999 spectra that HD 179821 was an F star in the 1990's.
 
%\subsection{Spherical Shells vs. discs}

We model the full spectral energy distribution of HD 179821 in order to fit photometric (optical, 2MASS, sub-millimeter, millimeter) and spectroscopic (ISO) data.  In particular, we compare SED model fits from both codes to investigate degeneracies between circumstellar dust geometries.  In Section 4, we present an inwardly truncated, flared disc model.  The inner wall absorbs radiation from the central source and marks the transition to the optically thick outer disc.  The circumstellar material in this case corresponds to a torus-like geometry.  In Section 5, we present a spherically symmetric shell model using the radiative transfer code DUSTY.  

\section{Inner Wall, Edge-on Disc Models}

The central star of HD 179821 is assumed to be located at 5 kpc with the following properties: $T_{eff}=7000$ K, $R_\star=287$ $R_\odot$, $M_\star = 8$ $M_\odot$.  For these parameters, we are implicitly assuming that the system is a post red supergiant.  However, since our model scales with distance, our modeling results could also apply to a post-AGB system at 1 kpc.  

We consider an optically-thick disc with an inner wall at radius $r=R_w$ and height $H_w$.  The wall marks the transition between the evacuated interior region and  the optically-thick, outer disc.  The wall is illuminated at normal incidence by the central object and is assumed to be uniform in the vertical direction.  The radial structure is solved according to \citealt{2005ApJ...621..461D}.  At wavelengths $>35$ $\mu$m, emission is dominated by the outer disc.  Here, we focus on fitting the ISO spectrum in the case of an edge-on disc and discuss fitting the sub-millimeter and millimeter wavelengths in Section 5.  

We assume spherical dust grains in the wall atmosphere of size distribution $n(a)\sim a^{-3.5}$ between minimum and maximum radii $a_{min}$ and $a_{max}$ (\citealt{1977ApJ...217..425M}).    Opacities are calculated using Mie theory (\citealt{Wiscombe}).

We consider several silicate compositions (\citealt{1999A&AS..136..405H}, \citealt{2003JQSRT..79..765J}).  For each composition, we vary two parameters not constrained a priori: height of the wall and disc inclination angle.  In addition, we vary $a_{min}$ and $a_{max}$ to obtain the best spectral fit, leaving the power-law index fixed.  

The shape of the spectrum between  $\sim 8-30$ $\mu$m allows us to rule out several silicate compositions.  The 10 $\mu m$ silicate feature is smooth and does does not show evidence of the substructures associated with crystalline components.  This most likely indicates that the silicates are glassy and amorphous.  Therefore, we can immediately disregard crystalline silicates of mean cosmic composition (Mg$_{0.5}$Fe$_{0.43}$Ca$_{0.03}$Al$_{0.04}$SiO$_3$; \citealt{1994A&A...292..641J}) as they contribute substructures not observed in the smooth silicate features.  Additionally, the glassy silicates from \citealt{1994A&A...292..641J}, do not fit the spectrum longward of 30 $\mu$m and hence, we rule them out.  Hot shell circumstellar silicates (\citealt{1992A&A...261..567O}) do not provide a reasonable fit to the ISO spectrum.  If instead we use cool shell circumstellar silicates (\citealt{1992A&A...261..567O}), we can fit the 10 $\mu$m feature but can not match the emission longward of 20 $\mu$m.

%This is consistent with the fact that the circumstellar material is oxygen-rich (\citealt{1999AJ....117.1834R}).

We obtain reasonable fits to the ISO spectrum of HD 179821 using glassy pyroxene (optical constants from \citealt{1997A&A...327..743H} and \citealt{1995A&A...300..503D}), glassy olivine (Mg$_{1}$Fe$_{1}$SiO$_4$; \citealt{1995A&A...300..503D}) and glassy bronzite (\citealt{1988A&A...198..223D}).  Fig. \ref{model1} presents our fits for glassy pyroxene for two different inclinations: $\mu\equiv cos(i)=0.25$ (top) and $\mu=0.45$ (bottom).  For both figures, the temperature, $T_0$, at the innermost radius of the wall is $128$ K (see \citealt{2005ApJ...621..461D}).  In addition, the minimum and maximum grain radii which best fit the ISO spectrum are given by $a_{min}=0.005$ $\mu$m and $a_{max}=1.0$ $\mu$m.  We have tried many different values of $a_{max}$ over our range of parameters and grains of $\sim1$ $\mu$m and larger are required for our best fits.  For the top model in Fig. \ref{model1}, the position of the wall is $R_w=3.7\times10^3$ AU with a height of $H_w=1.73\times10^3$ AU.  We find good agreement up to 20 $\mu$m at which point our model undershoots the observed SED.

%Photometry points are represented by diamonds and our synthetic photosphere is the thin solid line.  The dark solid line is the SED produced by our wall model with glassy pyroxene for two different inclinations: $\mu\equiv cos(i)=0.25$ (left) and $\mu=0.45$ (right).  For both figures, the temperature, $T_0$, at the innermost radius of the wall is $128$ K (see \citealt{2005ApJ...621..461D}).  In addition, the minimum and maximum grain radii which best fit the ISO spectrum are given by $a_{min}=0.005$ $\mu$m and $a_{max}=1.0$ $\mu$m.  We have tried many different values of $a_{max}$ over our range of parameters and grains of $\sim1$ $\mu$m and larger are required for our best fits.  For the left model in Fig. \ref{model1}, the position of the wall is $R_w=3.7\times10^3$ AU with a height of $H_w=1.73\times10^3$ AU.  We find good agreement up to 20 $\mu$m at which point our model undershoots the observed spectra.  

If $\mu$ is increased (i.e. tilted more toward face-on), as in the bottom figure, it is possible to increase the emission slightly at 20 $\mu$m however the flux at 18 $\mu$m starts to deviate from the ISO spectrum.  The position of the wall in this model is the same, but a slightly shorter wall is required; $H_w=1.69\times10^3$ AU.  

Fig. \ref{model2}, compares disc models obtained for two different silicate compositions: glassy olivine (top) and glassy bronzite (bottom).  For the top figure, the best fit requires a slightly hotter wall, $T(r_i)=131$ K, located at $R_w=6.3\times10^3$ AU.  The wall height is given as $H_w = 3.4\times10^3$ AU.  For olivine, we require $a_{min}=0.005$ $\mu$m and $a_{max}=5.0$ $\mu$m for our best fit, however, while we find good overall agreement between 2 and 45 $\mu$m, our model produces excess emission between 15 and 18 $\mu$m.  For bronzite, we find that $T_0=128$ K with $R_w=6.7\times10^3$ AU and $H_w=3.7\times10^3$ AU.  For this model, the grain sizes are $a_{min}=0.005$ $\mu$m and $a_{max}=2.0$ $\mu$m.  

Using all three silicate compositions, we find that $R_w=5.2\pm1.5\times10^3$ AU and $H_w=2.7\pm1.0\times10^3$ AU with the wall dust temperature ranging from $128$ to $131$ K.  With a limited library of dust opacity data, we cannot determine the exact composition of the wall, other than to suggest that it could be a mixture of our suggested materials or perhaps could be fit by other amorphous silicates.  For instance, \citealt{1992ApJ...392L..75J} fit the 10 and 20 $\mu$m features from their UKIRT CGS3 spectra using a colder ($\sim90 - 95$ K) mixture of olivine and magnetite.  

To summarize, we found satisfactory fits to the ISO spectrum with the disc code but could not fit photometry longward of 40 $\mu$m.  We aim to compare modeling results from spherical and disc codes in regimes which might produce overlap.  Thus, we study the shell model in the next section.

%We model the spectral energy distribution using a state-of-the-art radiative transfer wall code developed and described by D'Alessio et al. (2005).  This code self-consistently solves the complete set of vertical disc structure equations including irradiation from the central source, viscous heating and accretion shocks on the central source surface.  We account for different degrees of dust mid-plane settling while also allowing for varying compositional structure.  In addition, disc gaps are included by modeling a sharp transitional wall between the optically thin inner region and optically thick outer disc.  Radiative transfer in the wall atmosphere is carried out accordingly (see model of CoKu Tau/4, GM Aur and DM Tau by Calvet et al. [2005]).  The particle size distribution, composition and spatial location of the dust, in addition to inclination, accretion rate and viscosity coefficient are varied (or input if known) to obtain the best fit to the entire SED, including (when available) 2MASS and IRAC broadband photometry, the IRS spectrum, and sub-millimeter and millimeter photometry.

\section{Spherical Shell Models}

We use the spherically symmetric, radiative transfer code DUSTY (\citealt{DUSTY}).  Given an incident radiation field and specified dusty region, the code self-consistenly calculates the emergent flux including dust scattering, absorption and emission in a spherical, non-rotating environment.  In addition, for radiatively driven winds, DUSTY computes wind structure by jointly solving the hydrodynamic equations.  The dust density profile within the shell is given as $\rho(r)\propto r^{-2}$, consistent with a constant mass loss rate.

DUSTY was previously used to model HD 179821.  \citealt{2002Ap&SS.281..751S} produced a best fit shell model with an inner radius of $r_i=8.7\times10^{3}$ AU and temperature $T(r_i)=130$ K.  However, it is difficult to tell how good a fit it is as they omit the ISO spectra in their model.  \citealt{2007A&A...462..637B} also modeled HD 179821 using DUSTY and obtained a slightly cooler shell ($T(r_i)=110$K) located at a farther distance ($r_i=12.0\times10^3$ AU).  However, neither investigation used the ISO spectrum to constrain their models.  We reproduced both previous DUSTY models.  However, when overlaid on the SED with the added constraint of the ISO spectrum, neither model proved satisfactory.  As the ISO spectrum provides the most stringent constraints on model parameters, our fit improves previous work.

For our spherical model, the temperature at the inner shell boundary ($T(r_i)=128$ K) was varied until the model infrared excess matched the ISO spectrum.  We fix the grain distribution as $n(a)\propto a^{-3.5}$ and vary the maximum, $a_{max}$,  and minimum, $a_{min}$, grain sizes.  The equilibrium temperature of a dust grain depends on size and composition.  Small grains superheat while large grains exhibit blackbody emission.  At each radius within the nebula, DUSTY assumes that the dust grains are at the same temperature.  This approximation is a generic feature of both codes used in this paper and warrants acknowledgment.  Additionally, both codes assume isotropic scattering.  Future work should incorporate anisotropic scattering in both geometries.

In addition, we vary the ratio of the outer to inner shell radius, $Y\equiv \frac{r_o}{r_i}$.  The central radiation source is assumed to be the same 7000 K model photosphere used previously in \S 4.  The optical depth through the shell at $0.55$ $\mu$m is also varied.  We opt for a radiatively driven wind and vary the composition of the grains.

Our best fit to the data yields an optical depth through the shell of  $\tau_V=1.95\pm0.03$.  In addition, we find a good fit using minimum and maximum grain sizes $a_{min}= 0.005$ $\mu$m and $a_{max}=0.25$ $\mu$m.  We find the ratio of the outer to inner shell radius produces similar results for $Y\equiv \frac{r_o}{r_i} = 15\pm2$.  

The best fit is obtained using a dust composition such that 95\% of the grains are a mixture of amorphous silicates with the other 5\% composed of iron oxide (17:1:1:1 ratio of interstellar silicates to glassy olivine to glassy pyroxene to wustite \citealt{1984ApJ...285...89D,1995A&A...300..503D,1995A&AS..112..143H}).  We have extensively varied the grain composition using our limited library and found this composition to yield the best result.  Wustite (FeO; \citealt{1995A&AS..112..143H}) is included as it provides a slight source of long wavelength emission between $40$ and $50$ $\mu m$ which could not be fit by extending the outer shell radius or increasing the maximum grain size.  However, even by including wustite, our model still has a slight deficit of emission in the $40-50$ $\mu$m region.  Wustite is featureless and the boost in emission is minor, thus we can not say with any certainty whether it is present in HD 179821.

Our spherical shell result is presented in Fig. \ref{Dusty1}.  The model spectrum (solid dark line) is plotted on the $A_v=2$ dereddened, $T=5750$ K spectrum.  Even though the central radiation source is a F-star ($T=7000$ K), an optical depth through the shell of $\tau_V=1.95$ obscures the photosphere so that it resembles a G-star ($T=5750$ K).  In particular the overall fit appears quite good in fitting data longward of 40 $\mu$m.  Both our disc and shell models produce similar results for the ISO spectrum.  Sub-millimeter and millimeter photometry highlights the main difference between a shell spectrum (small range of temperatures) and a disc spectrum (broad range of temperatures).  

The best shell fit yields an inner radius of $r_i=9.8\times10^3$ AU which is in excellent agreement with mid-IR images and polarimetry observations (\citealt{2001MNRAS.328..301G}).  The placement of the inner shell radius is also in agreement with \citealt{2002Ap&SS.281..751S}, however we obtain an outer shell radius, $r_0$, that is a factor of 6 smaller.  This result may be more in line with polarization images in the near-IR which show a circularly symmetric reflection nebula with a diameter of $15''$, a factor of $\sim 0.75$ times our outer radius (\citealt{1993ASPC...45..151K}).  \citealt{2007A&A...462..637B} assumed to a $n(a)\propto a^{-6}$ dust distribution power law when fitting HD 179821 with DUSTY.  While they arrived at a slightly cooler, closer shell, we found a satisfactory fit without appealing to steeper distributions by using a hotter central star.  It is useful to mention that we reproduced both previous model fits (\citealt{2002Ap&SS.281..751S,2007A&A...462..637B}).  When the ISO spectrum was included, neither of those two previous models matched observations.  In general, with a $T_{eff}=5750$ K central source, we could not reproduce the infrared excess.  To match the shape and peak of the infrared excess required a hotter, obscured central source ($T_{eff}=7000$ K).

Additionally from our fit, the dust temperature in the outer shell is $T(r_o)=41$ K.  DUSTY also provides an estimate of the mass outflow rate although this value should be treated with caution as it has an inherent uncertainty of $\sim30$\%.  Varying the gravitational correction by 50\% produces no discernible effect on the spectrum and is responsible for the uncertainty in the mass outflow rate (see \citealt{DUSTY} for more detail).  Our derived outflow rate is $\dot{M}=2\times10^{-4}$ $M_\odot$ $yr^{-1}$ again in good agreement with \citealt{2002Ap&SS.281..751S} and a factor of $\sim5$ less then \citealt{2007A&A...462..637B}.  We summarize results below.

\section{Summary}
We have modeled the broad-band spectral energy distribution of HD 179821 using two distinct radiative transfer codes: one corresponding to a spherical shell geometry with $\rho\left(r\right)\propto r^{-2}$, and one to a disc geometry with $\rho\left(r\right)\propto r^{{-3}/{2}}$.  Under these assumptions, both codes provided equally good fits for similar dust compositions and size distributions shortward of $40$ $\mu$m.  However, longward of $40$ $\mu$m, only the spherical shell model reproduces the observations.  The radial dust density profile and corresponding range of temperatures present in the disc provide excess emission above the spherical shell solution and cannot fit the data.  A wavelength of $\sim$ 40 $\mu$m marks the boundary for discerning the model fits.  

It should be noted that the density profiles need not uniquely indicate a particular geometry.  If the densities deviate from the anticipated theoretical values of $\rho\left(r\right)\propto r^{-2}$ (shell), $\rho\left(r\right)\propto r^{-3/2}$ (disc), then the infrared excess is likely coupled to the density distribution rather than the geometry.  The effect of changing the density profile in both codes should be investigated before comparative spectral modeling is established as a reliable method for distinguishing geometric differences.

%{\textit {For unresolved objects such as point sources, data longward of $\sim40$ $\mu$m must be obtained before one can distinguish geometric differences from spectral modeling.}}

Based on our detailed spectral modeling, we conclude that it is likely that the nebula around HD 179821 is a radiatively driven shell.  The dust is mainly composed of small ($a_{max}=0.25$ $\mu$m) amorphous, glassy silicates.  Interior to the shell is an evacuated region in addition to the central radiation source.  The central star is most likely a $T=7000$ K star obscured such that it appears as a $T=5750$ K star (optical depth of $\tau_V\sim1.95$ through the spherical shell).  This may aid in explaining previous conflicting spectral classifications.  The dimensions of our shell agree well with previous observations and provide a good fit to the ISO spectrum.  

\begin{figure}
\leavevmode
\begin{center}
\includegraphics[height=3.0in]{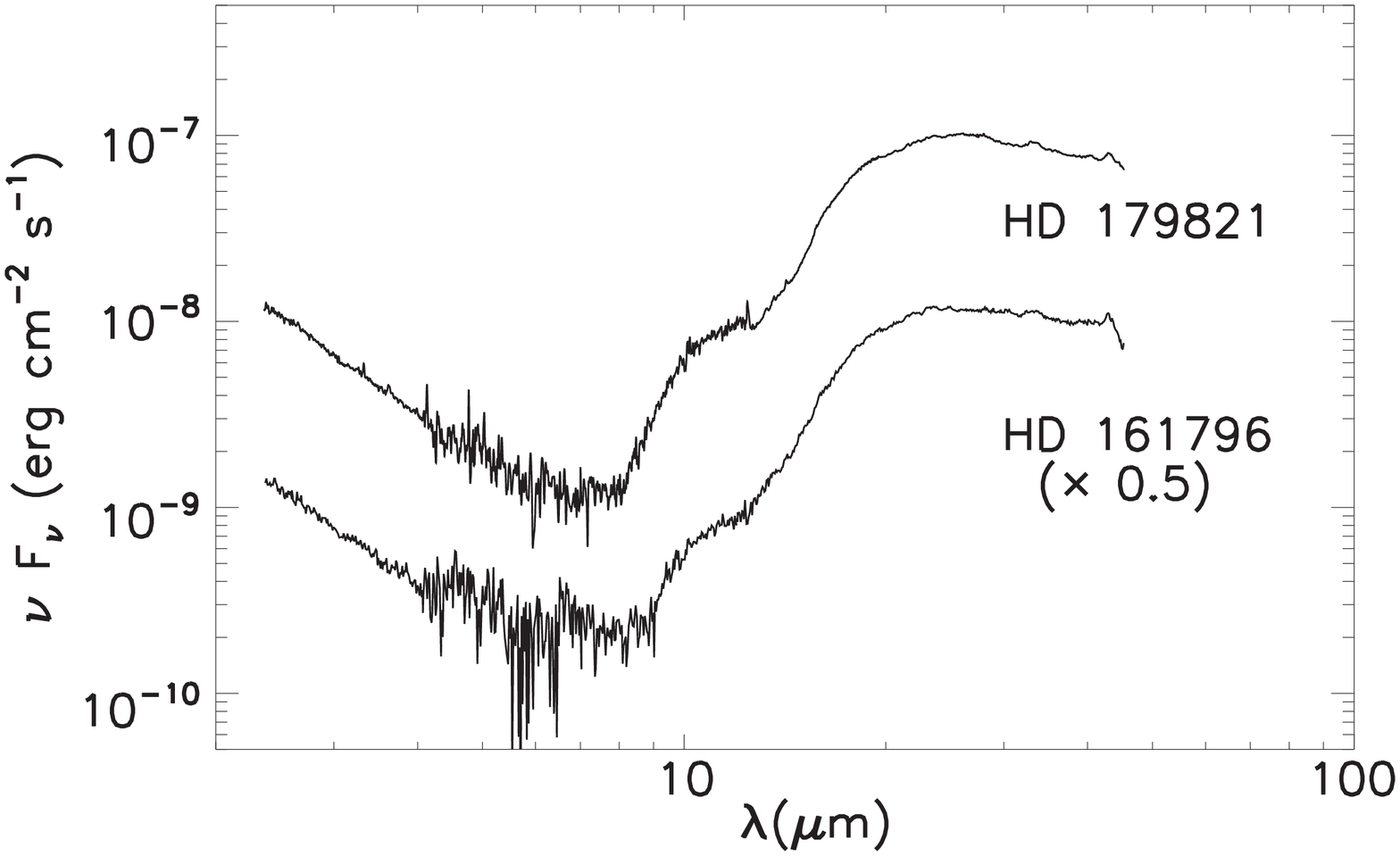}
\vspace{0.5in}
\hspace{0.0in}
\includegraphics[height=3.0in]{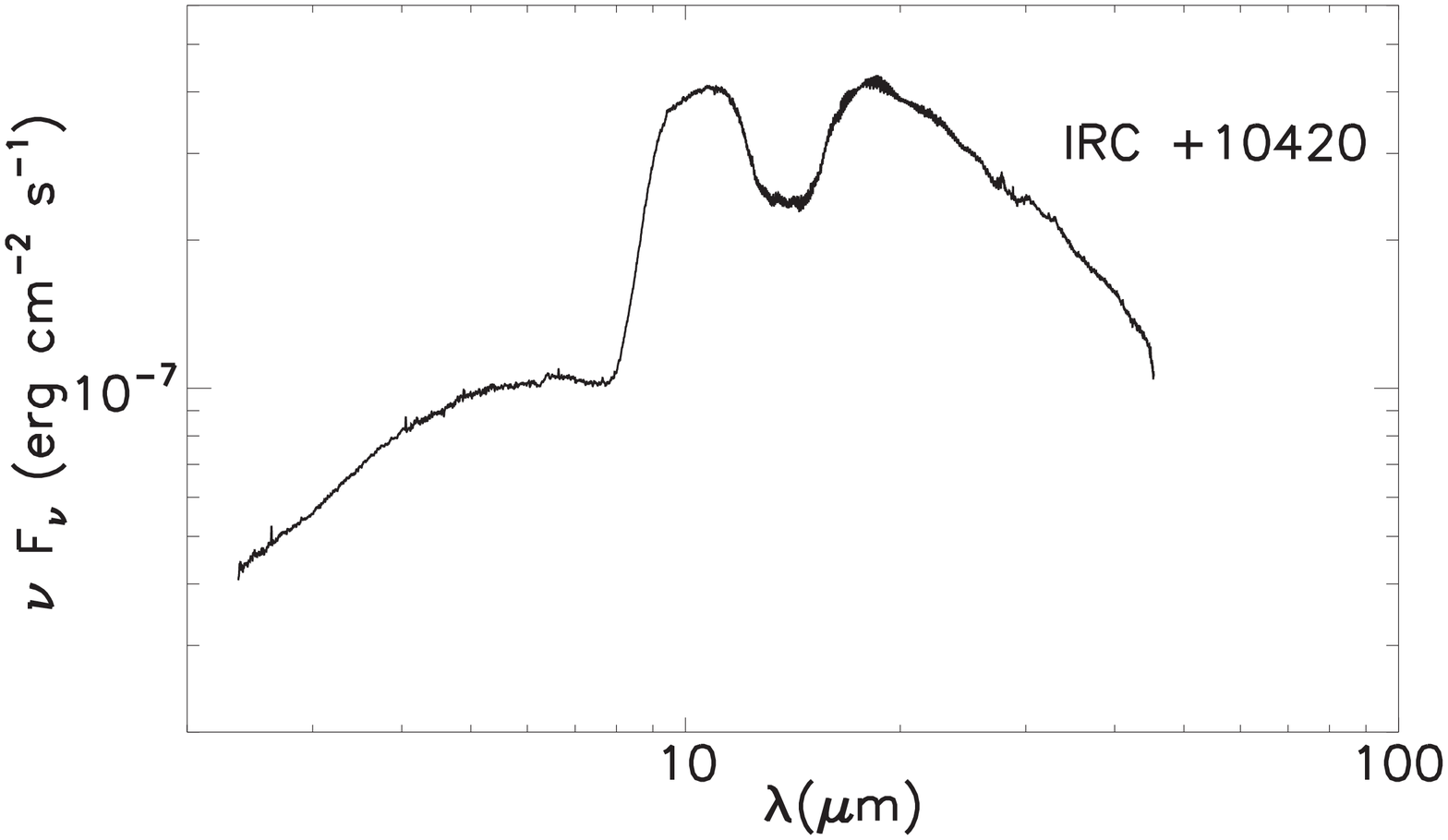}
\caption{Top: ISO SWS spectra of HD 179821 and post-AGB object HD 161796.  In both objects, there is a steep rise between the 10 and 20 $\mu$m features possibly indicating a transition region to the optically thick outer wall or shell.  HD 179821 was de-reddened using $A_V=2$ while HD 161796 was corrected using $A_V=1.2$. Bottom: ISO spectrum of the post-RSG IRC +10420. \label{geometry}}
\end{center}
\end{figure}

\begin{figure}
\leavevmode
\begin{center}
\includegraphics[height=3.0in]{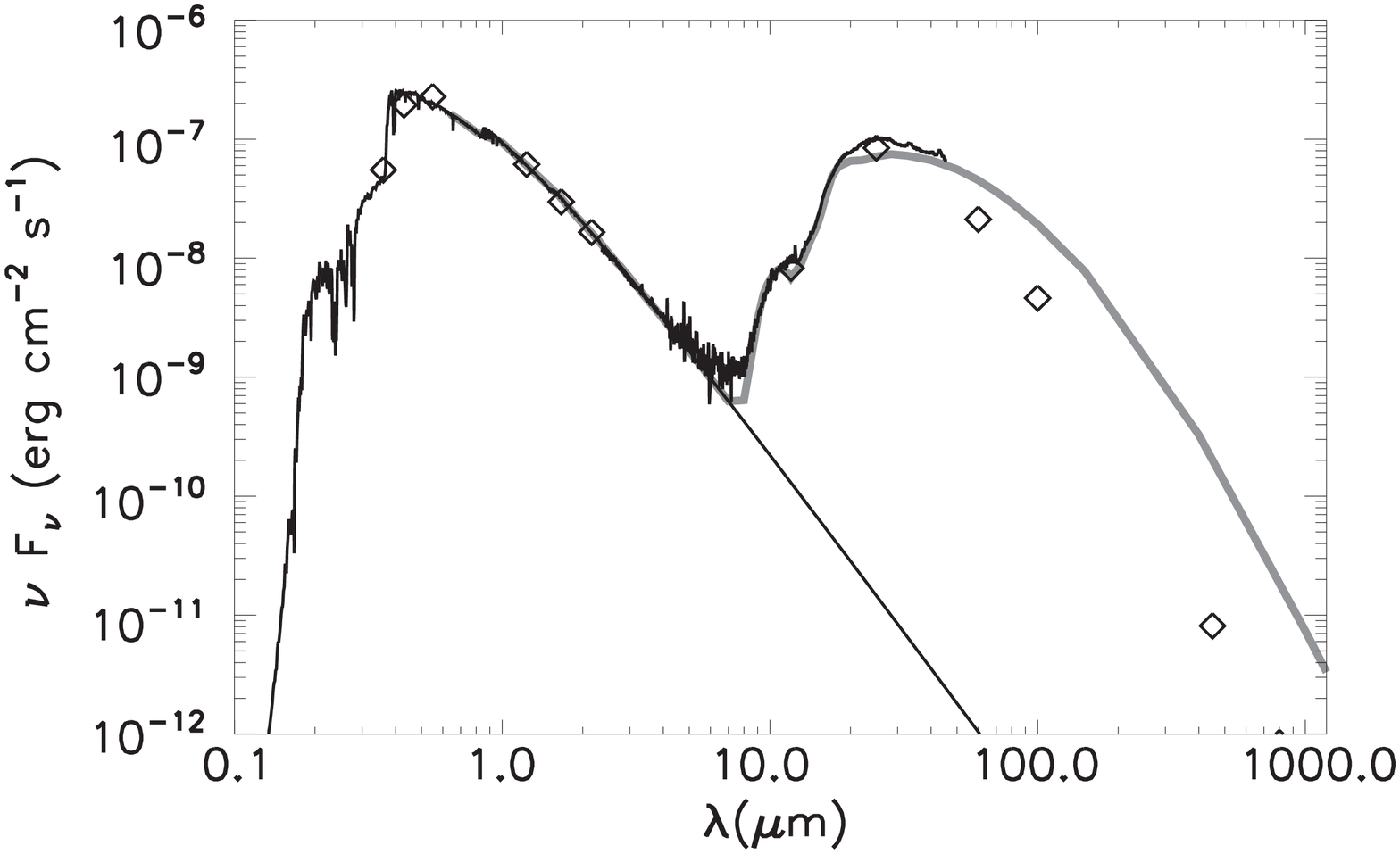}
\vspace{0.5in}
\hspace{0.0in}
\includegraphics[height=3.0in]{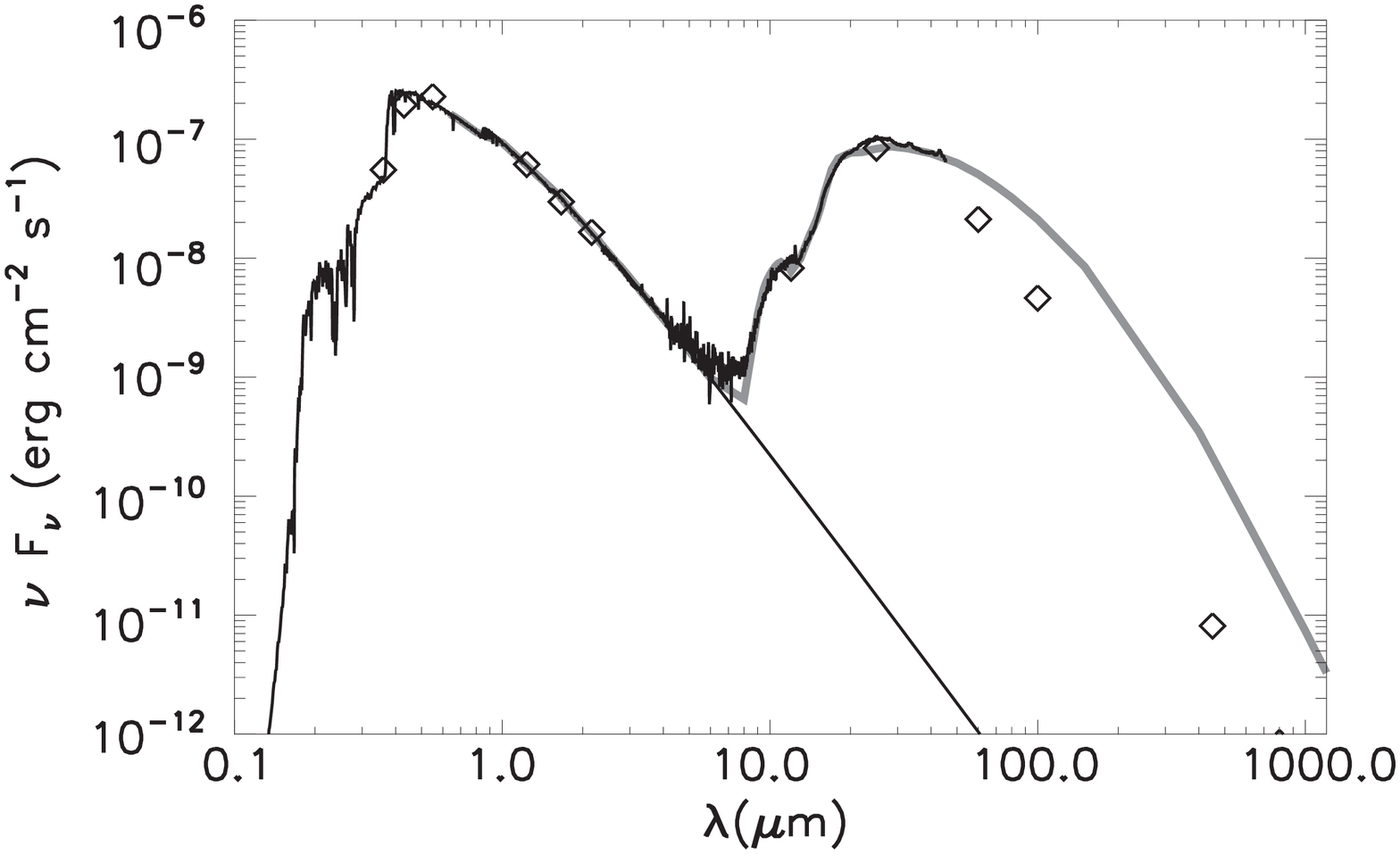}
\caption{Our wall models.  The top figure corresponds to $\mu=0.25$ while the bottom corresponds to $\mu=0.45$. \label{model1}}
\end{center}
\end{figure}

\begin{figure}
\leavevmode
\begin{center}
\includegraphics[height=3.0in]{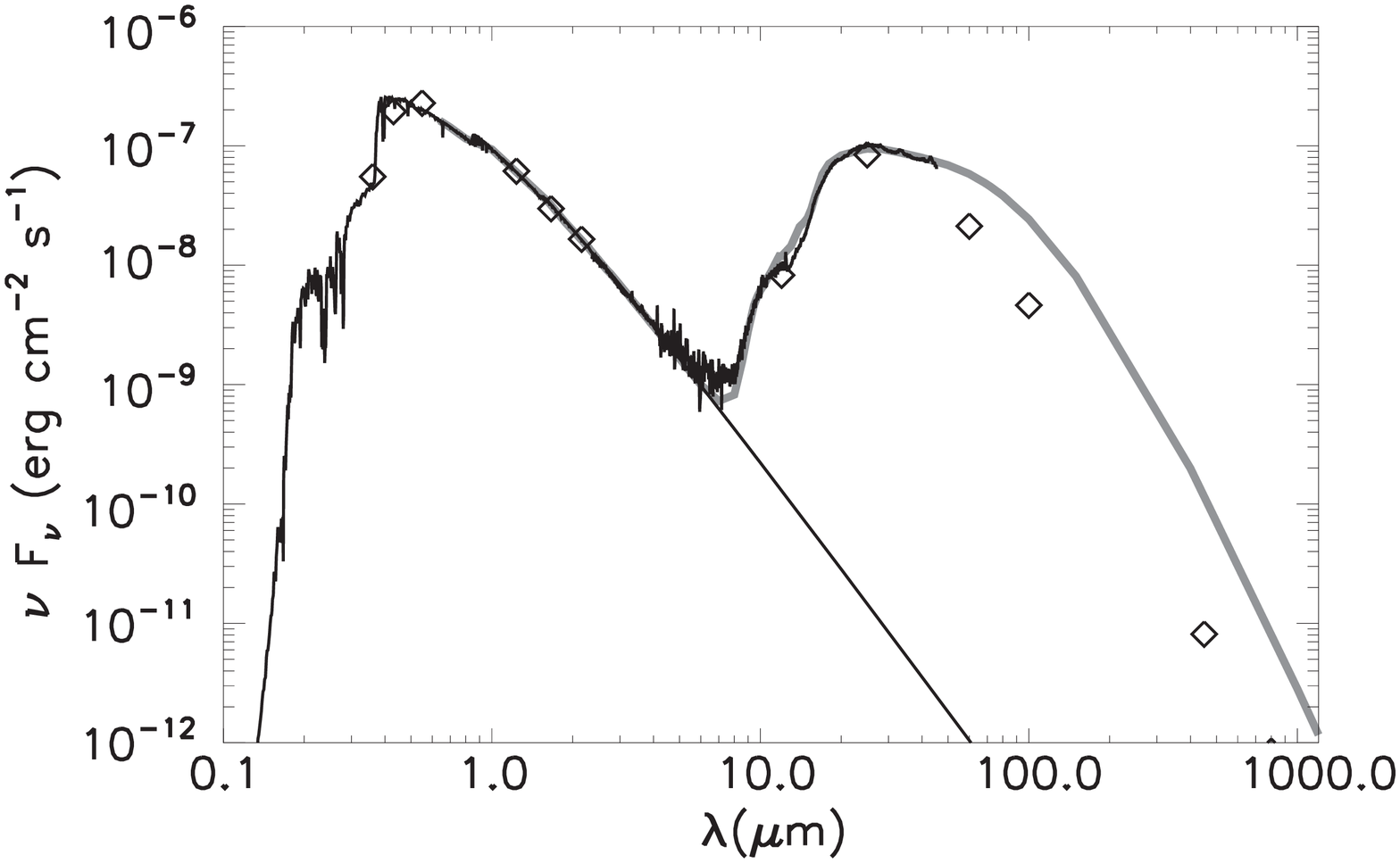}
\vspace{0.5in}
\hspace{0.0in}
\includegraphics[height=3.0in]{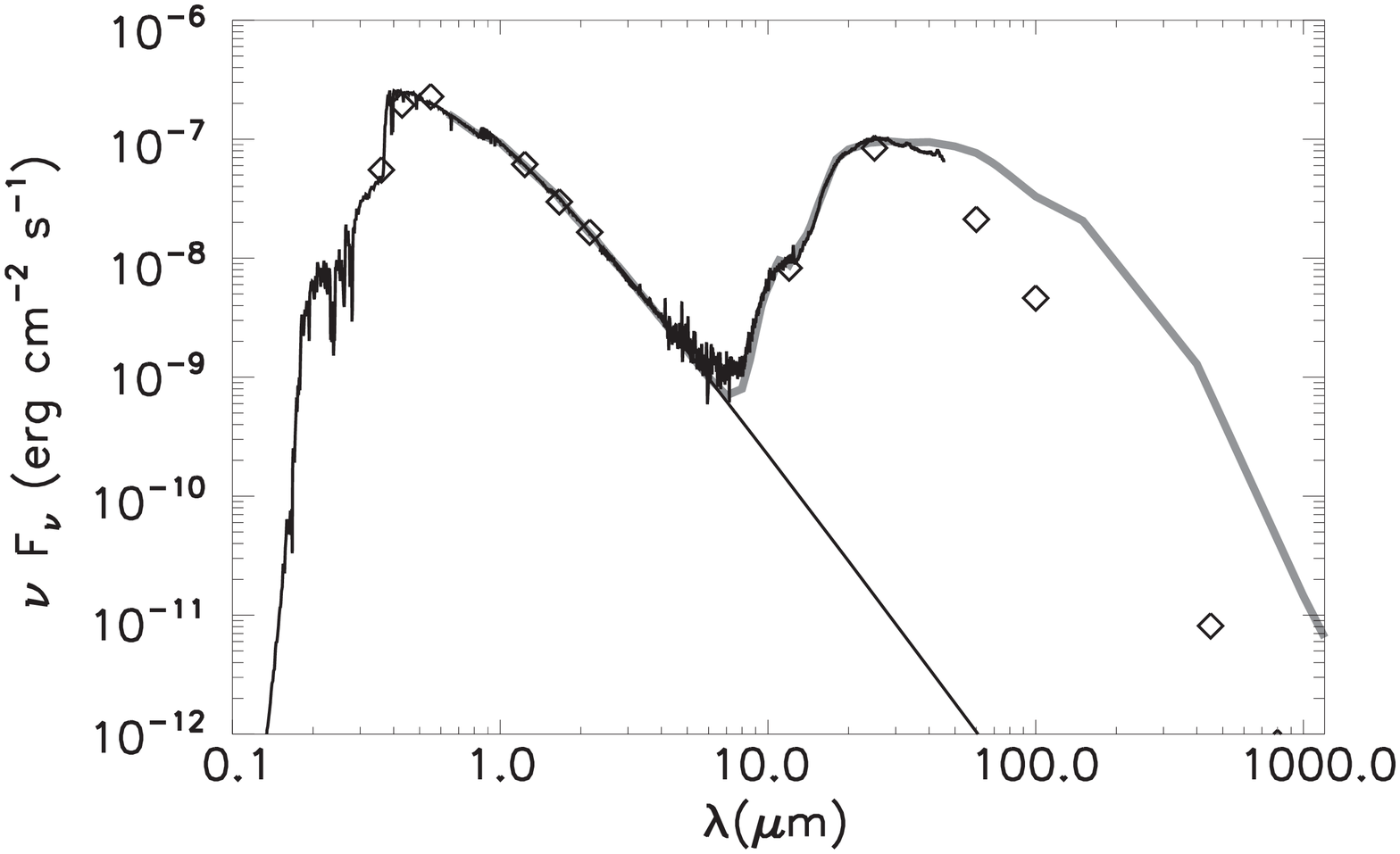}
\caption{Another fit to the wall model for glassy olivine (top) and glassy bronzite (bottom).  The inclination in both figures is $\mu=0.25$. \label{model2}}
\end{center}
\end{figure}

\begin{figure}
\plotone{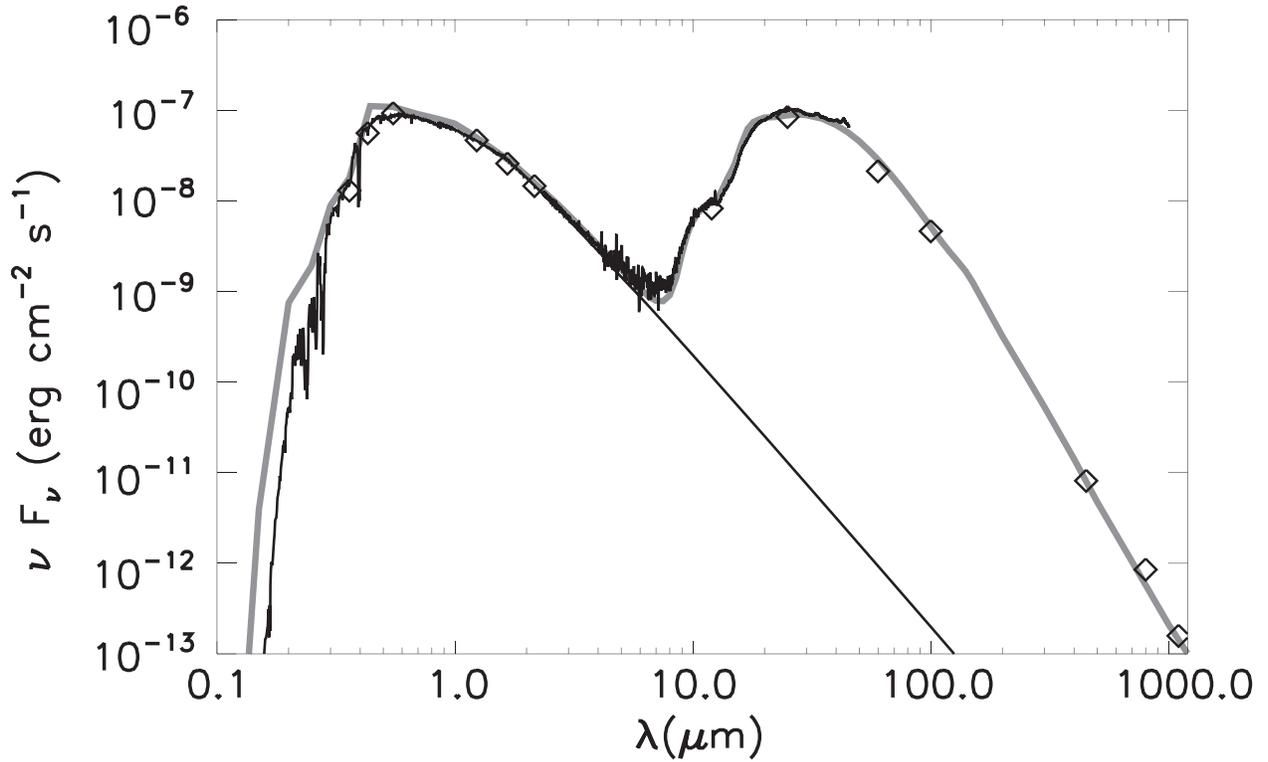}
\caption{Our spherical shell model (dark line).  The dust temperature at the inner and outer radii are 128 K and 41 K.  Even though the central radiation source is a $T=7000$ K photosphere, the detached shell reddens the central source so that it resembles a $T=5750$ K photosphere (thin line).  The silicate features are fit using a mixture of glassy, amorphous silicates with a small component of FeO. \label{Dusty1}}
\end{figure}

We compared the ISO spectrum to that of HD 161796, a confirmed post-AGB object (e.g. Fig. \ref{geometry}).  In particular, the ISO spectra and broad-band SED look remarkably similar.  Because our results scale with distance, our spectral modeling does not provide insight into whether HD 179821 is a post-AGB or post-RSG.  However, the similarity of the ISO spectra suggests that HD 161796 and HD 179821 experienced a very similar mass-loss history.

\acknowledgements{We thank Tim Gledhill for insightful comments which led to an improved manuscript.  JTN acknowledges financial support of a Horton Fellowship from the Laboratory for Laser Energetics through the U. S. Department of Energy and HST grant AR-10972. In addition, we thank P. D'Alessio and N. Calvet for use of their disc codes.  IM acknowledges support from NSF grant ASST-0406823 and NASA grant NNG04GM12G.  EGB acknowledges support from NSF grants AST-0406799, AST-0406823, and NASA grant ATP04-0000-0016 (NNG05GH61G).  AF acknowledges support from JPL Spitzer grant 1278931, NSF grant AST-0507519 and DOE grant DE-F03-02NA00057.}

{}

%\begin{figure}
%\leavevmode
%\begin{center}
%\includegraphics[height=3.0in]{HD179821_dered_Av_2_Teff_5750_logg_1.5-scaled.eps}
%\vspace{0.5in}
%\hspace{0.0in}
%\includegraphics[height=3.0in]{HD179821_dered_Av_3_Teff_7000_logg_0.5-scaled.eps}
%\caption{SEDs of HD 179821.  Top: The photosphere is a Kurcurz, $T_{eff}=5750$ K star.  The data was dereddened using $A_v=2$.  The photospheric $\nu F_\nu$ emerging from the stellar surface was multiplied by $1.75\times{10^{-18}}$ to fit the photometric data.  Bottom: The photosphere is a Kurcurz, $T_{eff}=7000$ K model.  The data was dereddened using $A_v=3$.  Here the photospheric $\nu F_\nu$ is scaled by $1.65\times{10^{-18}}$.  Photometric and spectral data are give in Tables \ref{photometry} and \ref{IRAS}. \label{Av2}}
%\end{center}
%\end{figure}

\newpage
\clearpage
\begin{footnotesize}

\begin{deluxetable}{lcccccccc}
\tablewidth{0pt}
\tablecaption{Photometric Data \label{photometry}}
\tablehead{
\colhead{\textsc{Object}}                       &
\colhead{\small{U}}                      &
\colhead{\small{B}}           &
\colhead{\small{V}}             &
%\colhead{\small{R}}               &
%\colhead{\small{I}}                       &
\colhead{\small{J}}                      &
\colhead{\small{H}}           &
\colhead{\small{Ks}}             &
%\colhead{\small{L}}               &
%\colhead{\small{M}}                       &
%\colhead{\small{N}}                      &
%\colhead{\small{Q}}           &
%\colhead{\small{S.T.}}             &
%\colhead{\small{$A_v$}}               
}
\startdata
%HD 161796 &7.82&7.54&7.08&6.67&6.42&5.83&5.98&6.02&5.83&...&...&-2.5&F3 Ib&0.1\\
HD 179821& 10.81 & 9.49 & 7.89 & 5.371 & 4.998 & 4.728\\
\enddata
\tablecomments{Magnitude summary for HD 179821.  UBV measurements are on the Johnson system and are reproduced from \citealt{1989ApJ...346..265H}.  JHKs measurements are from the 2MASS point source catalogue (\citealt{2MASS}).
}
\end{deluxetable}
\end{footnotesize}

\begin{footnotesize}
\begin{deluxetable}{cccccccc}
\tablewidth{0pt}
\tablecaption{Colour Corrected IRAS fluxes, Submillimeter Data and ISO Spectra for HD 179821 \label{IRAS}}
\tablehead{
\colhead{\small{F$_{12}$$_{\mu m}$}}                      &
\colhead{\small{F$_{25}$$_{\mu m}$}}           &
\colhead{\small{F$_{60}$$_{\mu m}$}}             &
\colhead{\small{F$_{100}$$_{\mu m}$}}               &
\colhead{\small{F$_{450}$$_{\mu m}$}}                       &
\colhead{\small{F$_{800}$$_{\mu m}$}}                      &
\colhead{\small{F$_{1100}$$_{\mu m}$}}       &
\colhead{\small{ISO TDT}}               
}
\startdata
32$\pm$ 3& 700 $\pm$ 50 & 430 $\pm$ 30 &160 $\pm$ 30&1.22 $\pm$ 0.12& 0.227 $\pm$ 0.017 & 0.058 $\pm$ 0.009&11301444
\enddata
\tablecomments{
All presented fluxes have units of Janskys.  Data is reproduced from \citealt{1994A&A...285..551V}.  ISO spectra is processed according to \citealt{2003ApJS..147..379S} where multiplicative corrections are used for all wavelengths.
}
\end{deluxetable}
\end{footnotesize}

\begin{footnotesize}

\begin{deluxetable}{clcccccccccc}
\tablewidth{0pt}
\tablecaption{Model Summary \label{modeltable}}
\tablehead{
\colhead{\textsc{Model}}							&
\colhead{\textsc{Composition}}                       &
\colhead{\small{$T_{eff}$}}                      &
\colhead{\small{$T_0$}}           &
\colhead{\small{$a_{min}$}}             &
%\colhead{\small{R}}               &
%\colhead{\small{I}}                       &
\colhead{\small{$a_{max}$}}                      &
\colhead{\small{$\mu$}}           &
\colhead{\small{$H_w$}}             &
\colhead{\small{$\tau_V$}}               &
\colhead{\small{$Y\equiv\frac{r_o}{r_i}$}}                       &
\colhead{\small{Fits $>$ 40 $\mu m$}}                      &
%\colhead{\small{Q}}           &
%\colhead{\small{S.T.}}             &
%\colhead{\small{$A_v$}}               
}
\startdata
%HD 161796 &7.82&7.54&7.08&6.67&6.42&5.83&5.98&6.02&5.83&...&...&-2.5&F3 Ib&0.1\\
\textsc{Disc}&&K & K &$\mu$m &$\mu$m & & AU& & &\\
\hline
Fig. \ref{model1} &Pyroxene & 7000 &128   &0.05 &1.0  & 0.25  & 1.73$\times10^3$  &- & -& No&\\
Fig. \ref{model1}&Pyroxene & 7000 &128   &0.05 &1.0  & 0.45  & 1.69$\times10^3$  &- & -& No &\\
Fig. \ref{model2}&Olivine & 7000 &131   &0.005 &5.0  & 0.25  & 3.40$\times10^3$  &- & -& No &\\
Fig. \ref{model2}&Bronzite & 7000 &128   &0.005  &2.0  & 0.25  & 3.70$\times10^3$ &- & -& No &\\
\hline
\textsc{Shell}&& &  & & & & & & &\\
\hline
Fig. \ref{Dusty1}&Mixture & 7000 &128   &0.005  &0.25  & -  & - &1.95 & 15& Yes &\\
\enddata
\tablecomments{Summary of our disc and spherical shell models.  
}
\end{deluxetable}
\end{footnotesize}


\clearpage
\begin{thebibliography}{}


\bibitem[Arkhipova et al.(2001)]{2001AstL...27..156A} Arkhipova, V.~P., 
Ikonnikova, N.~P., Noskova, R.~I., Sokol, G.~V., \& Shugarov, S.~Y.\ 2001, 
Astronomy Letters, 27, 156
\bibitem[Balick \& Frank(2002)]{2002ARA&A..40..439B} Balick, B., \& Frank, A.\ 2002, \araa, 40, 439
\bibitem[Bidelman(1981)]{1981AJ.....86..553B} Bidelman, W.~P.\ 1981, \aj, 
86, 553 
\bibitem[Buemi et al.(2007)]{2007A&A...462..637B} Buemi, C.~S., Umana, G., 
Trigilio, C., \& Leto, P.\ 2007, \aap, 462, 637
\bibitem[Bujarrabal et al.(1992)]{1992A&A...257..701B} Bujarrabal, V., 
Alcolea, J., \& Planesas, P.\ 1992, \aap, 257, 701
\bibitem[Bujarrabal et al.(2001)]{2001A&A...377..868B} Bujarrabal, V., 
Castro-Carrizo, A., Alcolea, J., \& S{\'a}nchez Contreras, C.\ 2001, \aap, 
377, 868
\bibitem[Buscombe(1984)]{1984mscs.book.....B} Buscombe, W.\ 1984, {\it MK Spectral Classifications. Sixth General Catalogue}, Evanston: 
Northwestern University, 1984
\bibitem[Calvet et al.(2005)]{2005ApJ...630L.185C} Calvet, N., et al.\ 
2005, \apjl, 630, L185
\bibitem[Castro-Carrizo et al.(2007)]{2007A&A...465..457C} Castro-Carrizo, 
A., Quintana-Lacaci, G., Bujarrabal, V., Neri, R., \& Alcolea, J.\ 2007, 
\aap, 465, 457
\bibitem[Charbonnel(1995)]{1995ApJ...453L..41C} Charbonnel, C.\ 1995, 
\apjl, 453, L41
\bibitem[D'Alessio et al.(2001)]{2001ApJ...553..321D} D'Alessio, P., 
Calvet, N., \& Hartmann, L.\ 2001, \apj, 553, 321
\bibitem[D'Alessio et al.(2005)]{2005ApJ...621..461D} D'Alessio, P., et 
al.\ 2005, \apj, 621, 461
\bibitem[Decin et al.(2006)]{2006A&A...456..549D} Decin, L., Hony, S., de 
Koter, A., Justtanont, K., Tielens, A.~G.~G.~M., \& Waters, L.~B.~F.~M.\ 
2006, \aap, 456, 549
\bibitem[De Marco et al.(2004)]{2004ApJ...602L..93D} De Marco, O., Bond, 
H.~E., Harmer, D., \& Fleming, A.~J.\ 2004, \apjl, 602, L93
\bibitem[De Marco et al.(2007)]{2007arXiv0709.1508D} De Marco, O., Wortel, 
S., Bond, H.~E., \& Harmer, D.\ 2007, ArXiv e-prints, 709, arXiv:0709.1508 
\bibitem[Dorschner et al.(1988)]{1988A&A...198..223D} Dorschner, J., 
Friedemann, C., Guertler, J., \& Henning, T.\ 1988, \aap, 198, 223 
\bibitem[Dorschner et al.(1995)]{1995A&A...300..503D} Dorschner, J., 
Begemann, B., Henning, T., Jaeger, C., \& Mutschke, H.\ 1995, \aap, 300, 
503 
\bibitem[Draine \& Lee(1984)]{1984ApJ...285...89D} Draine, B.~T., \& Lee, 
H.~M.\ 1984, \apj, 285, 89
\bibitem[Edgar et al.(2007)]{2007arXiv0709.2292E} Edgar, R.~G., Nordhaus, 
J., Blackman, E., \& Frank, A.\ 2007, ArXiv e-prints, 709, arXiv:0709.2292
\bibitem[Gledhill et al.(2001)]{2001MNRAS.328..301G} Gledhill, T.~M., 
Yates, J.~A., \& Richards, A.~M.~S.\ 2001, \mnras, 328, 301
\bibitem[Hawkins et al.(1995)]{1995ApJ...452..314H} Hawkins, G.~W., 
Skinner, C.~J., Meixner, M.~M., Jernigan, J.~G., Arens, J.~F., Keto, E., \& 
Graham, J.~R.\ 1995, \apj, 452, 314
\bibitem[Henning et al.(1995)]{1995A&AS..112..143H} Henning, T., Begemann, 
B., Mutschke, H., \& Dorschner, J.\ 1995, \aaps, 112, 143
\bibitem[Henning \& Mutschke(1997)]{1997A&A...327..743H} Henning, T., \& 
Mutschke, H.\ 1997, \aap, 327, 743
\bibitem[Henning et al.(1999)]{1999A&AS..136..405H} Henning, T., Il'In, 
V.~B., Krivova, N.~A., Michel, B., \& Voshchinnikov, N.~V.\ 1999, \aaps, 
136, 405
\bibitem[Hrivnak et al.(1989)]{1989ApJ...346..265H} Hrivnak, B.~J., Kwok, 
S., \& Volk, K.~M.\ 1989, \apj, 346, 265 
\bibitem[Hrivnak et al.(1994)]{1994ApJ...420..783H} Hrivnak, B.~J., Kwok, 
S., \& Geballe, T.~R.\ 1994, \apj, 420, 783
\bibitem[Jaeger et al.(1994)]{1994A&A...292..641J} Jaeger, C., Mutschke, 
H., Begemann, B., Dorschner, J., \& Henning, T.\ 1994, \aap, 292, 641
\bibitem[Ivezi{\'c} et al.(1999)]{DUSTY} Ivezi{\'c}, Z., Nenokva, M., \& Elitzur, M. 1999, User Manual for DUSTY, University of Kentucky Internal Report
\bibitem[J{\"a}ger et al.(2003)]{2003JQSRT..79..765J} J{\"a}ger, C., Il'in, 
V.~B., Henning, T., Mutschke, H., Fabian, D., Semenov, D., \& 
Voshchinnikov, N.\ 2003, Journal of Quantitative Spectroscopy and Radiative 
Transfer, 79, 765
\bibitem[Josselin \& L{\`e}bre(2001)]{2001A&A...367..826J} Josselin, E., \& 
L{\`e}bre, A.\ 2001, \aap, 367, 826
\bibitem[Jura \& Werner(1999)]{1999ApJ...525L.113J} Jura, M., \& Werner, 
M.~W.\ 1999, \apjl, 525, L113
\bibitem[Justtanont et al.(1992)]{1992ApJ...392L..75J} Justtanont, K., 
Barlow, M.~J., Skinner, C.~J., \& Tielens, A.~G.~G.~M.\ 1992, \apjl, 392, 
L75 
\bibitem[Kastner \& Weintraub(1993)]{1993ASPC...45..151K} Kastner, J.~H., 
\& Weintraub, D.\ 1993, Luminous High-Latitude Stars, 45, 151
\bibitem[Kastner \& Weintraub(1995)]{1995ApJ...452..833K} Kastner, J.~H., 
\& Weintraub, D.~A.\ 1995, \apj, 452, 833
\bibitem[Keenan(1983)]{1983BICDS..24...19K} Keenan, P.~C.\ 1983, Bulletin 
d'Information du Centre de Donnees Stellaires, 24, 19
\bibitem[Kraemer et al.(2002)]{2002ApJS..140..389K} Kraemer, K.~E., Sloan, 
G.~C., Price, S.~D., \& Walker, H.~J.\ 2002, \apjs, 140, 389
\bibitem[Kurucz(1993)]{Kurucz} Kurucz, R. L. 1993, ATLAS9, CD-ROM 13 (Cambridge, MA: Smithsonian Astrophysical Observatory)
\bibitem[Likkel(1989)]{1989ApJ...344..350L} Likkel, L.\ 1989, \apj, 344, 350 
\bibitem[Mathis et al.(1977)]{1977ApJ...217..425M} Mathis, J.~S., Rumpl, 
W., \& Nordsieck, K.~H.\ 1977, \apj, 217, 425
\bibitem[Moe \& De Marco(2006)]{2006ApJ...650..916M} Moe, M., \& De Marco, 
O.\ 2006, \apj, 650, 916 
\bibitem[Nenkova et al.(2000)]{2000ASPC..196...77N} Nenkova, M., 
Ivezi{\'c}, {\v Z}., \& Elitzur, M.\ 2000, Thermal Emission Spectroscopy 
and Analysis of Dust, discs, and Regoliths, 196, 77
\bibitem[Nordhaus \& Blackman(2006)]{JN2006} Nordhaus, J., \& 
Blackman, E.~G.\ 2006, \mnras, 370, 2004 
 \bibitem[Nordhaus et al.(2007)]{JN2007} Nordhaus, J., 
Blackman, E.~G., \& Frank, A.\ 2007, \mnras, 376, 599
\bibitem[Nordhaus \& Blackman(2007)]{2007arXiv0707.3792N} Nordhaus, J., \& 
Blackman, E.~G.\ 2007, ArXiv e-prints, 707, arXiv:0707.3792
\bibitem[Oudmaijer et al. (2008)]{2008arXiv0801.2315O} Oudmaijer, R., 
Davies, B., de Wit, W.-J., \& Patel, M.\ 2008, ArXiv e-prints, 801, 
arXiv:0801.2315
\bibitem[Ossenkopf et al.(1992)]{1992A&A...261..567O} Ossenkopf, V., 
Henning, T., \& Mathis, J.~S.\ 1992, \aap, 261, 567
\bibitem[Perryman et al.(1997)]{1997A&A...323L..49P} Perryman, M.~A.~C., et 
al.\ 1997, \aap, 323, L49
\bibitem[Quintana-Lacaci et 
al.(2007)]{2007A&A...471..551Q} Quintana-Lacaci, G., Bujarrabal, V., Castro-Carrizo, A., \& Alcolea, J.\ 2007, \aap, 471, 551 
\bibitem[Reddy \& Hrivnak(1999)]{1999AJ....117.1834R} Reddy, B.~E., \& 
Hrivnak, B.~J.\ 1999, \aj, 117, 1834
\bibitem[Reyes-Ruiz \& L{\'o}pez(1999)]{1999ApJ...524..952R} Reyes-Ruiz, 
M., \& L{\'o}pez, J.~A.\ 1999, \apj, 524, 952
\bibitem[Scheffler \& Elsaesser(1987)]{1987pgim.book.....S} Scheffler, H., 
\& Elsaesser, H.\ 1987, {\it Physics of the Galaxy and Interstellar Medium}, Berlin and New York, Springer-Verlag, 1987, 503 
p.
\bibitem[Skrutskie et al.(2006)]{2MASS} Skrutskie, M.~F., et 
al.\ 2006, \aj, 131, 1163
\bibitem[Sloan et al.(2003)]{2003ApJS..147..379S} Sloan, G.~C., Kraemer, 
K.~E., Price, S.~D., \& Shipman, R.~F.\ 2003, \apjs, 147, 379
\bibitem[Soker(2006)]{2006ApJ...645L..57S} Soker, N.\ 2006, \apjl, 645, L57
\bibitem[Surendiranath et al.(2002)]{2002Ap&SS.281..751S} Surendiranath, 
R., Parthasarathy, M., \& Varghese, B.~A.\ 2002, \apss, 281, 751
\bibitem[Th{\'e}venin et al.(2000)]{2000A&A...359..138T} Th{\'e}venin, F., 
Parthasarathy, M., \& Jasniewicz, G.\ 2000, \aap, 359, 138
%\bibitem[van der Veen et al. (1989)]{1989A&A...226..108V} van der Veen, 
%W.~E.~C.~J., Habing, H.~J., \& Geballe, T.~R.\ 1989, \aap, 226, 108
\bibitem[Ueta et al.(2000)]{2000ApJ...528..861U} Ueta, T., Meixner, M., \& 
Bobrowsky, M.\ 2000, \apj, 528, 861
\bibitem[van der Veen et al.(1994)]{1994A&A...285..551V} van der Veen, 
W.~E.~C.~J., Waters, L.~B.~F.~M., Trams, N.~R., \& Matthews, H.~E.\ 1994, 
\aap, 285, 551
\bibitem[Wasserburg et al.(1995)]{1995ApJ...447L..37W} Wasserburg, G.~J., 
Boothroyd, A.~I., \& Sackmann, I.-J.\ 1995, \apjl, 447, L37
\bibitem[Wiscombe(1979)]{Wiscombe} Wiscombe, W. J. 1979, Mie Scattering Calculations: Advances in Technique and Fast, Vector-Speed Computer Codes, NCAR T/N-40+STR (Boulder: NCAR)
\bibitem[Zacs et al.(1996)]{1996MNRAS.282.1171Z} Zacs, L., Klochkova, 
V.~G., Panchuk, V.~E., \& Spelmanis, R.\ 1996, \mnras, 282, 1171
\bibitem[Ziurys et al.(2007)]{2007Natur.447.1094Z} Ziurys, L.~M., Milam, 
S.~N., Apponi, A.~J., \& Woolf, N.~J.\ 2007, \nat, 447, 1094
\bibitem[Zuckerman \& Dyck(1986)]{1986ApJ...311..345Z} Zuckerman, B., \& 
Dyck, H.~M.\ 1986, \apj, 311, 345

\end{thebibliography}
\end{document}